\documentclass[12pt]{article}

\begin{document}

\begin{center}
{\bf{\large
A Map of the Nanoworld:\\
Sizing up the Science, Politics, and Business of the Infinitesimal
}}

\bigskip

{\em Debashish Munshi,$^a$ Priya Kurian,$^b$ Robert V. Bartlett,$^c$ Akhlesh Lakhtakia$^{d,e}$}

\bigskip

$^a$Department of Management Communication,
Waikato Management School,\\
University of Waikato,
Private Bag 3105,
Hamilton, New Zealand\\

\bigskip

$^b$Department of Political Science and Public Policy,\\
University of Waikato,
Private Bag 3105,
Hamilton, New Zealand\\

\bigskip

$^c$Department of Political Science,\\
Purdue University,
100 N. University Street,
West Lafayette, IN 47907-2098, USA\\

\bigskip

$^d$Department of Engineering Science and Mechanics,\\
Pennsylvania State University,
University Park, PA 16802-6812, USA\\

\bigskip

$^e$Department of Physics,\\
Imperial College,
London SW7 2AZ, UK\\

\end{center}
 
\noindent{\bf Abstract:}
Mapping out the eight main nodes of nanotechnology discourse that have emerged in the past decade, we explore how various scientific, social, and ethical islands of discussion have developed, been recognized, and are being continually renegotiated. We do so by (1) identifying the ways in which scientists, policy makers, entrepreneurs, educators, and environmental groups have drawn boundaries on issues relating to nanotechnology; (2) describing concisely the perspectives from which these boundaries are drawn; and (3) exploring how boundaries on nanotechnology are marked and negotiated by various nodes of nanotechnology discourse. \\

\newpage

\section{ Introduction}

Nanotechnology is the art and science of making materials, devices, and systems with very small but very precise architectures that are invisible to optical microscopes. Can infinitesimal multi-molecular physical particles be technologically efficacious and yet socio-ethically acceptable? In this paper, we map the main nodes of nanotechnology discourse that have emerged in recent years.  We explore how various scientific, social, and ethical islands of discussion have developed, been recognized, and are being continually renegotiated. An understanding of the distinctive characteristics of these distinct islands, their rates and directions of development, and the impact of any commerce among them will better allow researchers and policymakers to evaluate complex scientific and socioethical issues pertaining to technoscience and the environmentÑnamely, what in the future may be permitted, encouraged, or prohibited economically, legally, technically, socially, or morally, by whom, and in which contexts.

Nanotechnology is on the verge of becoming a ``gigaideology'' [1] but most discussions on this subject are not multidimensional, multidisciplinary, or fully open. Some discourse nodes focus on an imagined {\em utopia} of a superworld where minuscule nanorobots will be ``able to eliminate cancer, infections, clogged arteries, and even old age'' [2], or of a dreamland of virtual reality [3, 4]. The discourse in other nodes revolves around the prospect of either a {\em dystopia} of an unstable world [5] or of the fury of uncontrolled self-replication that can run amok in the world [6].

To our knowledge, no attempt has been made as yet to survey and describe these islands of discussion, to examine their interactions and degrees of isolation, or, indeed, to point out any commonalities among them.  The lack of consensus in the views on nanotechnology, in fact, has much to do with the multiple conceptions of the notion of boundaries.  Boundaries are not simple demarcations along scientist-public lines, inasmuch as there are as many conflicting conceptions among nanoscientists themselves as there are among journalists, business leaders, and social-humanistic researchers.

Our aims in this paper are: (1) to identify how scientists, policymakers, entrepreneurs, educators, and environmentalists have drawn boundaries on issues relating to nanotechnology; (2) to describe concisely the perspectives from which these boundaries are drawn; and (3) to explore how boundaries on nanotechnology are marked and negotiated by various nodes of nanotechnology discourse.

The process of demarcating boundaries starts with the very definition of nanotechnology. While the US definition is that at least one dimension of a nanoparticle or the relevant length scale of an exploited phenomenon must lie between 1 and 100 nanometers (nm) long for a product or a process to be called nanotechnology [7, 8], British thinkers insist that the nanotechnological range is between 0.2 nm to 100 nm [9]. The magnitude of the difference in these two positions can be gauged by the fact that 1 nm is far less than a mere speck in the eye. Chop an inch-long piece of thread into 25 equal pieces and then chop each of those pieces into a million equal pieces. Those tiny pieces would be about 1 nm long.  A human hair is about 80,000 nm in diameter, a red blood cell is about 7000 nm in diameter, globular proteins are 6 nm in diameter, quantum dots are 1 to 2 nm in diameter, and the crucial gate oxide layers in metal-oxide-semiconductor devices that are the workhorses of integrated electronics are nowadays 2 nm thick. These are all manufacturable dimensions today. But a hydrogen atom is about 0.1 nm in diameter, which means that the sub-nm length scales have traditionally belonged to chemistry. Is nanotechnology then the same, even partially, as chemistry? The answer to that question depends on one's perspectives. Notwithstanding the prominence of chemists in nanotechnology, many nano-technoscientists distance themselves from chemists. 

Nanotechnology may use chemistry, but is not merely chemistry. It is also physics, electrical engineering, mechanical engineering, computational science, materials science, and more. The delineation of the distinctive features of nanotechnology is not an easy task yet, as John DiLoreto of the American Chemistry Council recently acknowledged [10]. In part, the problems in exact definition come from the amorphous nature of nanotechology, which made a British panel argue in favor of the term ``nanotechnologies'' rather than ``nanotechnology'' [9].

Examining the literature produced in the technoscientific disciplines that nanotechnology spans as well as on the socioethical ramifications of nanotechnology in the near future, we find that ignorance of some or all key aspects of nanotechnology is widespread. This feature sets the current societal discussions on nanotechnology apart from the discussions on genetically modified plants and animals, inasmuch as the techniques used to analyze the latter are not fruitful for the former.

\section{Nodes of nanotechnology discourse}

Our analysis of the literature on nanotechnology reveals the following eight nodes of societal discussion on nanotechnology: 

(1) Technoscientists, especially those either working on or supervising some 
nanotechnological application who, almost invariably, tend to glorify nanotechnology;
 
(2) Leaders of business and industry who want to cash in on the projected benefits by developing a market for nanotechnology-driven products; 

(3) Official or quasi-official bodies that generate a significant amount of literature; 

(4) Social science and humanities researchers who tend to focus on the social, economic, political, legal, religious, philosophical, and ethical implications of nanotechnology; 

(5) Fiction writers with imaginative scenarios, both utopian and dystopian; 

(6) Political activists, particularly those with an environmental worldview, who tend to extend to nanotechnology the issues long  raised by them with regard to biotechnology; 

(7) Journalists and popular science writers who report on current events, perspectives, and funding regimes relating to the field; and 

(8) John Q. and Jane D. Public, who are yet to significantly grapple with or discuss nanotechnology in any depth.

\noindent	We elaborate on each of the eight nodes in turn.

\subsection{Node 1: Technoscientists}

Research at universities is driven by funding from government agencies as well as from industries. This funding was responsible for the spectacular growth of US universities during and after the Second World War, and the lessons learnt therefrom are now being used by technoscientific managers as well as policymakers in Europe, Japan, China, and India. Furthermore, industrial research is normally predicated on financial success. In this environment, not surprisingly, entrepreneurial technoscientists---whether at universities,  industries, or government agencies---have learnt to align their research efforts with the latest terms in vogue. While most researchers are motivated by funding possibilities for their research areas, not a small percentage of them may have hopes of monetary benefits from commercialization of their research products.

In consequence, it seems that nanotechnology is suddenly everywhere in technoscientific circles. Many advances that were inevitable due to normal improvements and advances in conventional R\&D are being attributed to nanotechnology, just because size-reduction techniques have brought one, two, or all three length dimensions into the neighborhood of 100 nm. Cosmetics containing ultrafine clays and oil particles feel much better on the skin; plastics reinforced with carbon nanofibers are likely much stronger yet lightweight; better lithographic techniques have densified the doping of semiconductor real estate in integrated chips; and so on. These normal improvements are not revolutionary; instead, they exemplify the so-called {\em incremental technology} [11] that has fueled much hype in the business world and continues to do so [12]. There is anecdotal evidence in technoscientific circles that those technoscientists who are either unable or reluctant to nanofy their research areas by mere size-reduction are contemptuous of this hype.

There is, of course, nanotechnology research going on that elicits much admiration from technoscientists. Carbon nanotubes, quantum dots, sculptured thin films, single-electron transistors, nanofluidic sensors, and biomimetic substances are all examplar products of {\em evolutionary nanotechnology} [1, 13]. None of these advances has yet had any significant presence in the marketplace, and all kinds of miraculous benefits are claimed for the entire world based on mere projections of laboratory experiments. Yet there is something exciting about every example of evolutionary nanotechnology, whether in the product or in the process or both, as it has features that cannot be explained simply by size-reduction.

The initial, and still unfulfilled, promise of nanotechnology is bolder than evolutionary nanotechnology. It is called {\em radical nanotechnology} [11]. Widely cited is Richard FeynmanÕs 1959 vision of ``manipulating and controlling things on a small scale'', so that the ``entire 24 volumes of the Encyclopedia Brittanics [be written] on the head of a pin'' [14]. Eric Drexler in 1981 translated that vision into the concept of molecular manufacturing [15], which has engendered apocalyptic visions of nanomachines running amok and producing what some have described as ``grey goo''  [16]. But Drexler has been at pains to point out ``that the easiest and the most efficient systems will not have the capabilities required for autonomous runaway manufacturing'' [17], and indeed technoscientific capabilities for the foreseeable future fall short of FeynmanÕs vision [18]. Yet, scenarios encompassing radical technology not only have created hype but also fear in the general population [16, 19]. Visions of ``robot[s] the size of a blood cell'' and ``nanobots[s] that could do surgery without leaving any visible scars'' [20] exemplify as-yet unwarranted hype.  Increasingly, technoscientific commentators are warning against hype, possibly hoping to avert public backlash against funding for nanotechnology and even other technoscientific research.

Most of the technoscientific literature is glib enough to not indicate possible failures. This is normal in technoscientific literature, because the emphasis is on publishing positive and upbeat results, whereas negative issues are considered as unnecessary distractions. There are at least 25 print journals and at least one virtual journal that are either wholly or substantially dedicated to nanotechnology today, in addition to a vast array of other scientific and technical journals that occasionally publish accounts of nanotechnology research. A large number of scientific conferences have a few sessions dedicated to nanotechnology, not to mention specialist conferences on nanotechnology. Major learned societies have formed Technical Groups, Working Groups, Subcommittees, etc., to manage the explosive growth in nanotechnology. Yet, the (British) Institute of Physics' {\em Nanotechnology} is perhaps the only technoscientific journal that has occasionally published articles not written by nanaotechnology researchers. Of these articles, there is only one on socioethical issues emanating from possible industrial and economic success in nanotechnology [21]. 

Skepticism about nanotechnology as a panacea has been rarely offered by technoscientists on economic grounds [22, 23]. Likewise, safety issues have been rarely raised by nano-technoscientists [24, 25, 26]. However, the 2004 report of the Royal Society of London and the Royal Academy of Engineering [9] greatly improves on two earlier US reports [8, 27] to raise warning flags about dangers to public health as well as to environment, and vehemently cautions against complacency as well as continued ignorance. The health risks of nanotechnology are expected to use the same biological pathways as many potential applications of nanotechnology in medicine [9]. Whether or not the US Food \& Drug Administration is justified in ``believ[ing] that the existing battery of pharmacotoxicity tests is probably adequate for most nanotechnology products that [FDA] will regulate'' [28] is a moot point, given the lack of serious research on the toxicity of nanomaterials [9], no doubt because of competition from conventional products that cost much less to fabricate [29].

\subsection{Node 2: Leaders of business and industry}

The world of business and industry has ushered in the age of nanotechnology with cautious optimism. In their 2003 book [30], Uldrich and Newberry set the agenda for a climate of enthusiasm. Within a decade, the book proclaims, nanotechnology will be at the core of products worth a trillion dollars in the US alone and will drive commercial applications in an entire range of industries from agriculture to manufacturing. Uldrich and Newberry's wonderland of miniature marvels conjures up images of not only unimaginable improvements to existing products and services but also the invention of entirely new materials and products that will revolutionize the way we live.

Yet industry leaders have so far warmed up to the idea of investing in enterprises seeking the improvement of existing products and not so much to the creation of fundamentally new materials. Claims of improvement are most common for the paint and the cosmetics industries, as well as for their user industries. Semiconductor industries are also beginning to claim the benefits of nanolithography for shrinking device sizes and increasing packing densities in integrated chips. Many of the claimed advances are not only real but also cost-effective; but these advances fall in the realm of incremental nanotechnology, which is far from revolutionary. Venture capitalists, seeking to invest large amounts of liquid cash for relatively quick profits, generally form partnerships with university researchers with an entrepreneurial bent.  Nano start-up companies are based on a key patent or two, and the capital supplied by the venture capitalists is then invested to turn the patents into marketable products.

According to an exhaustive report [31] published by Lux Research, Inc., a U.S-based advisory firm that looks into the business dimensions of nanotechnology, the top start-up companies adopting nanotechnologies that have begun generating revenues are in the fields of specialty chemicals, pharmaceuticals, and semiconductor capital equipment. The report anticipates that activity in the nanotech sphere will soon shift from basic research to applications development as over 1500 start-ups worldwide develop.

The general aura of optimism is reflected in the number of major corporations that have decided to adopt nanotechnology. About ``63 per cent of the 30 companies comprising the Dow Jones Average (Dow) are currently funding R\&D in nanotechnology'' [31]. A major financial management company, Merrill Lynch, in fact, now publishes a `Nanotech Index' to help investors keep track of companies that deal with nanotechnology. ``We believe nanotechnology could be the next growth innovation,'' Merrill LynchÕs global technology strategist, Steven Milunovich, stated in a press release issued in April 2004 [32].

This optimism is, however, tempered by several notes of caution. As Baker and Aston [12] point out, Òinvestors, torn between an alluring new market and the fear of a dot-com-like bubble, are struggling to get a grip on exactly what nano means for them''. For many investors the promise of nanotechnology looks real enough for them to be interested in, but what keeps them back is a coherent translation of the scientific jargon behind much of the research being carried out in laboratories. This is where the complexities of boundaries surface again. The demarcation of boundaries is facilitated by composite sets of ``claims, activities, and institutional structures that define and protect knowledge practices'' [33] and in the case of nanotechnology, the knowledge practices of technoscientists are not yet easily readable on business spreadsheets. The \emph{boundary demarcation problem} is embedded in the discourses on nanotechnology that reflect conflicting desires to protect the unquestioned authority of scientific progress on the one hand (see Node 1) and to defend the basic profit-oriented instincts of business.

Business instincts are honed on experience but investors have not had much information in easily understood terms yet. The introduction to a 2005 international conference on nanotechnology and business held in Brussels stated [34] that despite nanotechnology being heralded as ``the next industrial revolution'', many businesses ``still don't understand the potential of the technology''. While encouraging businesses to look to this emerging field, this introduction described the field as complex and went on state that ``there is a vital need to understand the concepts and the pitfalls, and to recognize when the right time to invest is and which applications are commercially viable''.

The companies that are making significant investments in nanotechnology are ones that already have years of experience in the technology sector. Not surprisingly, ``BASF, Dow Chemical, DuPont, General Electric, Hewlett-Packard, IBM, and NEC hold most of the nanotech patentsÓ [35]. As indicated earlier, most of these companies are involved in incremental nanotechnology but hold out for a molecular revolution that will change the face of business.

\subsection{ Node 3: Official or quasi-official bodies}

A significant ``officialÓ literature has been generated by government agencies, international governmental organizations, and government-supported science and technology academies. This activity was undoubtedly generated as nanotechnology is ``widely seen as having huge potential [for] \dots many areas of research and application,'' [9, p. vii]  and is ``attracting investments from [g]overnments and from businesses'' [9, p. vii]. Furthermore, nanotechnology ``may raise new challenges in the safety, regulatory or ethical domains that will require societal debate.'' [9, p. vii]

The US government took the initiative early in September 1998 when it regularized a forum for inter-agency discussions that began in 1996 as the Interagency Working Group on Nanotechnology. In August 1999, after catalyzing and nurturing numerous workshops  and studies, this group submitted a draft plan for federal action. President Clinton announced the National Nanotechnology Initiative (NNI) in 2000. The NNI is a R\&D program to coordinate inter-agency efforts to realize the full potential of nanotechnology, to facilitate technology transfer to fuel the national economy, and to educational resources as well as a skilled workforce. Realizing that nanotechnology is a transformative social force, the NNI  coordinators regularly broadcast its activities to the general public on {\sf www.nano.gov}.

Societal issues were the focus of a workshop conducted by the US National Science Foundation in September 2000 [8]. In addition to the economic and industrial pathways for nanotechnology research and development, the conferees also examined the societal changes that nanotechnology could engender. Possible impacts on the environment, energy generation and consumption, water purification and desalination, agricultural yields, space exploration, national security, and the free-market economy were discussed. An adequate assessment of societal impacts may be possible only after several decades, because there are bound to be unintended or second-order consequences of nanotechnology, although the near-term emergence of radical nanotechnology was dismissed. The inclusion of social scientists in the NNI right from the beginning was recommended, as also the education of future social scientists on nanotechnology and its human dimensions.

The British government in 2003 commissioned the Royal Society and the Royal Academy of Engineering to define nanotechnology and to assess its current status and its future prospects, as well as to identify and predict socioethical challenges that nanotechnology could engender. After extensive consultation with technoscientists, policymakers, and others in July 2004, the British panel released a comprehensive report [9] that maps the opportunities of nanotechnology. The report did not cover socioethical uncertainties in much detail, possibly due to the dearth of data. As a result, the voices of technoscientists (who constitute the membership of the two academies) were more prominent in the report. Some social scientists did provide input to the Working Group of the panel, but the general lack of societal debate on the social implications of nanotechnology [36] left the social aspects underrepresented.

Among the critical issues raised by the British panel was the absence of standards for the nanotechnology workplace, both in terms of environmental impacts as well as with the possible risks of either deliberate or accidental introduction of nanoparticles in human beings and organisms.  Nanometrology is particularly important to the issue of workplace standards. The American Society for Testing of Materials International (ASTM Int'l) has set up Committee E56 on Nanotechnology to develop ``consensus standards'' [37]. Working with experts from six major partners---the Institute of Electrical and Electronics Engineers, the American Society of Mechanical Engineers, the international offices of the US National Science Foundation, the Americal Institute of Chemical Engineering, JapanÕs National Institute of Advanced Industrial Science and Technology, and Semiconductor  Equipment and Materials International---Committee E56 aims to develop an ``approved terminology document'' to streamline discussions in and between governments, industries, and academic research institutions [37].

The Japanese government takes its obligations to industry very seriously. It identified nanotechnology as one of four ``priority areas'' in its Second Science and Technology Plan [38]. A national committee for terminology and standards was established in late 2004. This committee is surveying Japanese organizations connected with nanotechnology and is also championing the creation of global standards  by the International Standards Organization.

Collectively, the many government and official reports constitute the node of this serious discourse that can be and is accessed across most of the other nodes, although most official reports themselves only acknowledge and explicitly draw upon the technoscientific and business discourse nodes (as well as other official reports).
 
\subsection{Node 4: Social science and humanities researchers}

Some early scholarly researchers from the social sciences and humanities have attempted to explore the social, economic, political, legal, religious, philosophical, and ethical implications of nanotechnology for human societies, but these researchers have not produced literatures yet, nor have they coalesced into functioning research communities.  This discourse node is still in a very early stage of development, which can be seen in its almost entirely outward focus---rarely do the scattered writings cite other published scholarly works in the humanities and social sciences, in part because even as late as 2005 there is still little to be cited.  Instead, occasional articles refer extensively to journalistic reports, reports from government agencies and official bodies, the self-promoting pronouncements of technoscientists and the business community, and the treatments of nanotechnology in fiction.

For example, only a skeletal legal scholarship addresses the implications of nanotechnology.  Although legal issues are frequently mentioned in non-legal periodicals and online forums, attention in law reviews and journals has been scarce. A pioneering overview article by Fiedler and Reynolds appeared in 1994 [39].  A few years later, Lin-Easton [40] offered a more specialized legal focus on the application of the precautionary principle of international environmental law, and Reynolds [41] explored environmental regulation of nanotechnology.  Developments in nanotechnology raise some obvious and not-so-obvious issues with regard to intellectual property rights, which were addressed by Newberger [42] and by Halluin and Westin [43]. Other articles, such as by Kerr and Bassi [44], appeared in law journals but consist more of political and ethical analysis than legal scholarship.  A general and superficial overview of the legal implications of nanotechnology across its projected stages is contained in Smith [45].  With the exception of Lin-Easton and Reynolds, these writers do not cite each other or other legal researchers.  Rather, the many and lengthy footnotes common to legal scholarship are filled with citations to works from the other nodes of discourse.

A similar pattern emerges in other areas of the humanities and social sciences.  Some  political scientists look at the political implications of nanotechnology, but tend primarily to cite literature from the technoscience, business, and official report nodes (e.g., Kay and Bosso [46]).  The most developed subliterature is that which focuses on the political implications of nanotechnology as it affects military security  [47]. Some ethicists have begun looking at nanotechnology [21, 48, 49], but much more common are ``lighthouse beacon'' articles by technoscientists and practitioners urgently calling for more serious attention to the profound ethical issues being raised by nanotechnology (e.g., [50]).  In the US, the National Science Foundation has funded initiatives to study the social implications of nanotechnology [8].  Much of the early literature derives directly from the promotional efforts of NSF officials [51, 52].  A few instances of social research not funded by NSF have also begun to appear [53].

Perhaps the most developed area in the social sciences, not surprisingly, is that of scientometrics, the measurement and analysis of science, which is often done through bibliometrics, the measurement and analysis of publications.  A significant literature exists that reports on research measuring and analyzing, for example, growth and trends [54], nanotechnology interdisciplinarity [55, 56, 57], patterns of research collaboration [55, 57], and patents [58].  This literature would seem to be highly relevant to nearly all other social science studies of nanotechnology, but so far it has been little cited outside of its own small research community.

Humanistic scholarship uses other discourse nodes as the raw material for analysis, and applies well-established humanistic theories and concepts, but the contributions are also few and scattered, do not speak to each other, and certainly do not build on each other.  Examples include Bendle's [59] and Elliott's [60] analyses of the transhuman or posthuman ideology underlying much of the discourse found in all the other nodes.  A more substantial scholarship consists of critical reviews of nanofiction and theoretical comparisons of it and various other nanotechnology narratives, part of the fiction node to which we now turn.

\subsection{ Node 5: Fiction writers }

Fiction writers from early on have explored the potentials of nanotechnology, raising questions that have in some instances then been taken up in other nodes.  Almost all of the emergent science fiction on nanotechnology has been based on the concerns of current science, even as it stretches any scientific consensus on what is plausible. And, as we have seen with the early science fiction of the 20th century, life can imitate art, and the blurring boundaries of fact and fiction are part of the development of a narrative in ``the construction of a new science and industry'' [61].

When Drexler presented a utopic vision of an age of nanotechnology in 1986 [62], while also identifying its dystopic underside---the ``grey goo'' problem when miniature ``assemblers'' replicate themselves endlessly---science fiction writers responded by charting the risks embedded in the topographies of the future. Perhaps the most popular work of this genre is Crichton's \emph{Prey} [63], illustrating the devastating consequences that result when cutting-edge technoscience joins hands with corporate greed and human fallibility. The novel paints a chilling picture of the convergence of nanotechnology, biotechnology, and computer science that leads to the creation of deadly ``swarms'' of self-replicating ``micro-robots''. Such a scenario may not be feasible given the state of nanotechnology today, but for Crichton, it seems inevitable that ``sometime in the twenty-first century, our self-deluded recklessness will collide with our growing technological power'' [63, p. x].

Another significant theme in some of the most popular science fiction envisions convergences in the fields of nanotechnology, biotechnology and computer science---not unlike the recent US government-funded initiative on NBIC (nanotechnology, biotechnology, information technology and cognitive science) [27]; see the discussion on Node 7. StephensonÕs {\em Diamond Age} is an influential novel that envisages the coming of a new age of nanotechnology, when ``matter compilers'' can make almost anything anyone could want [64]. 

These literatures, in turn, have given rise to a critical review literature. In a sweeping review of the scientific and science fiction writings on nanotechnology, Milburn pointed to the blurring of boundaries between the two despite (indeed, precisely because of) attempts by scientists to distance themselves from the ``negative associations of science fiction'' [65]. The imagined futures of nanotechnology conjured up by nanoscientists, which attract billions of dollars for R\&D from government and industry, are in fact coterminous with science fiction, embodying Baudrillard's notion of hyperreality [65].  So thin is the line between fact and fiction that ``we might almost say that reality is jealous of fiction, that the real is jealous of the image \dots . It is a kind of duel between them\dots'' [66, p. 28]. 

The Baudrillardian hyperreal concept of nanotechnology pictured in fiction, computer games, and films is more than just science or hard technology---it has evolved into a distinct culture [61]. In this culture, there is no distinction between what is human and what is non-human. If, as Fukuyama has argued Òthe most significant threat posed by contemporary biotechnology is the possibility that it will alter human nature and thereby move us into a `posthuman' stage of history'' [67, p. 7]  present-day nano-narratives already reveal our posthuman futures.

In addressing the latent perils of biotechnology, Fukuyama wrote that ``human nature shapes and constrains the possible kinds of political regimes, so a technology powerful enough to reshape what we are will have possibly malign consequences for liberal democracy and the nature of politics itself'' [67, p. 7]. This is already evident in the  Jonathan Demme film, \emph{The Manchurian Candidate} (2004) wherein nanotechnology is used to re-jig the central nervous system of a key player in a political tussle. Just as Alduous Huxley's book \emph{Brave New World} provided an advance peep into the world of biotechnology, the newer works of fiction with nano-narratives are providing the template for conceptualizing the posthuman future of nanotechnology. 

\subsection{Node 6: Political activists}

Activists, many with an environmentalist worldview, are wary of the race to the posthuman world. McKibben abhors the idea of giving up ``our citizenship in the land of the finite, which is the place that humans have known, and trade it for a passport to the infinite'' [5, p. 224]. Furthermore, he says that the children of the world should not have to contemplate being launched ``into a future without bounds, where meaning may evaporate'' or to ``live always in the future and never in the now where humans have always dwelt'' [5, p. 224].

In his 2003 book [5], McKibben has argued that society should turn its back on nanotechnology and other complex new technologies which threaten ``not just our survival, but our identity'' [68]. He is one of the growing numbers of members of action groups, think tanks, social movements, and churches that have become involved in evaluating the implications of nanotechnology for the environment, health, human rights, and global justice [69].

BritainÕs Prince Charles too has helped raise the salience of nanotechnology by his reported ``qualms'' about its possible impacts, which caused uproar among British scientists and the government who said he was distorting the debate [70]. Yet, as Goldsmith pointed out, prior to the princeÕs involvement, ``there was no debate. And\dots there were no signs that any such debate was about to be launched, despite the gold-rush excitement fuelling nanotechnologyÕs phenomenal expansion'' [71]. Although both McKibben and Prince Charles have been dismissed as naysayers by many technoscientists, more grudging respect is forthcoming for ETC (the Action Group on Erosion, Technology and Concentration), a Canada-based activist group, which has produced a series of influential reports on the social implications of nanotechnology [72, 73, 74]. It has urged caution in using nanotechnology, pointing to the possibility of  ``microorganisms \dots manipulated through nanotechnology to take over the function of machines but that begin reproducing out of control'', and has called for a moratorium on its development given evidence of the potential toxicity of nanoparticles [75]. The call for a moratorium was echoed by Greenpeace International in July 2003 [76, 77].

The website of ETC states that [78]:\begin{quote}
While nanotechnology offers opportunities for society, it also involves profound social and environmental risks, not only because it is an enabling technology to the biotech industry, but also because it involves atomic manipulation and will make possible the fusing of the biological world and the mechanical. There is a critical need to evaluate the social implications of all nanotechnologies; in the meantime, the ETC group believes that a moratorium should be placed on research involving molecular self-assembly and self-replication.\end{quote}
In its most recent report on nanotechnology, \emph{NanoGeoPolitics} [74], ETC has provided a critical overview of three different nanotechnology governance approaches that it says are emerging: ``(1) Optimists---`technology is good'---Full speed ahead (with `responsible' drivers at the wheel); (2) Realists---`technology is neutral'---Invite a few of the passengers to suggest alternative routes (the `upstream' approach); (3) Sceptics---`technology is political'---Get out the map and let everyone decide if they want to take a trip and if car, bike or bus is the best way to go'' [74, p. 7].

Activists---who have been politicized by policy debates over genetic modification of organisms---contribute in-depth reports, opinion pieces, and polemics to  periodicals and mainstream media outlets such as the \emph{New York Times} as well as to their own websites. Magazines such as \emph{The Ecologist} now run a regular column entitled 
\emph{Nanowatch} that keeps a close and critical watch on new developments in nanotechnology. Writing for this column on ``nano-pollution'', Thomas [79] described research where fish exposed to water containing nanotechnologyÕs 
``miracle molecules''---fullerenes, affectionately called ``buckyballsÓ by 
enthusiasts---suffered severe brain damage in just 48 hours. Thomas pointed out that this was the tenth report since the first published in 1997 
that warned of toxicity of nano-materials. The Ecologist has also published articles more broadly on the implications and risks of converging technologies, labelled variously as NBIC by the US government (see Node 5); GNR---genetics, nanotech and robotic---by Bill Joy of Sun Microsystems; and GRAIN---genetics, robotics, artificial intelligence and nanotech---by corporate environmental consultant Douglas Mulhall [80]. Thomas [80] quoted Ray Kurzweil of MIT as describing this convergence as ``the singularity---the point at which our technologies become the driving force in human evolution'' so much so that the ``world will be transformed beyond recognition'' [60]. Such convergences, hailed by techno-optimists such as Kurzweil, are viewed with considerable concern by anti-nanotechnology activists, some predicting the creation of destructive ``green goo of uncontrollable life forms'' arising from the linking of biotechnology and nanotechnology [19]. Others warn of the possibility in the near future of a ``nano surveillance society'' in the context of the ``war on terrorism'' [81]. 

Recognising that nanotechnology offers both potential benefits and risks, many of them yet unanticipated, some activists have called for the application of the precautionary principle as a way of managing nanotechnology [82]. Montague wrote that the insurance industry has expressed concern about the environmental and health hazards of nano-particles, and quotes a report by Swiss Re, a large reinsurance company that states, ``The Precautionary Principle demands the proactive introduction of protective measures in the face of possible risks, which science at present---in the absence of knowledge---can neither confirm nor reject'' [82]. A recent report indicates that other insurance companies are joining hands with academic scientists to evaluate the risks posed by manufactured nanomaterials [26].

	The varied activist reports and websites devoted to nanotechnology do monitor and respond to developments in Nodes 1 and 3, but with limited impact as yet on policy. It is primarily when their concerns get magnified through attention from the mainstream popular press that we see  some acknowledgement from the technoscientists and government research-funding bodies. It is to the discourses of the popular press that we now turn.

\subsection{Node 7: Science journalists and popular science writers}

	Negotiating the techno-utopian and techno-dystopian fault lines are professional science journalists and popular science writers. Such writers, who report on and analyze current cutting edge developments in science and technology, have produced a lively node of discussion about nanotechnology for more than a decade.  Some of these writers have sought to glorify the world that they see nanotechnology creating.  Others have suggested that public funding for nanotechnology be stopped or radically curtailed and governmental oversight be clamped on industrial efforts until the significant social, legal and ethical ramifications of nanotechnology have been mapped out; some argue for turning the clock back because no good will come out by tinkering with the fundamental mechanisms of nature.

In terms of sheer volume, much of the writing in this category comprises of short reports on current developments in nanotechnology; the opening of new nanotechnology centres, such as the one at Purdue University and the MIT-Army nanotechnology centre in 2002;  a plan by the European Commission and the US National Science Foundation to actively cooperate in research; and the great outpouring of funding for nanotechnology research and projects from governments and the private sector. Also to be found is reporting on nano-related events, such as the soaring futuristic visions of institutions including the Foresight Institute, whose founders believe that nanotechnology will give people ``mastery over matter'' [83]. In some ways, this node echoes the discourses of Nodes 1, 2, and 3 but there is very little, if any, exchange with Node 4.

Much less common is critical journalism that looks at the current nanohype with any degree of scepticism. Writing in the \emph{Wall Street Journal}, Gomes [84] urged caution in the face of venture capitalists working the Silicon ValleyÕs ``hype machine\dots gearing up to sell you a second-hand trend.'' The popular press has also reported on critiques of nanotechnology extended by groups such as the Canada-based ETC, which urge applying the precautionary principle to new technologies [85]. Indeed, the dangers of nanotechnology have been flagged by even technophiles with impeccable technological credentials such as Bill Joy and even the Foresight Institute that has called for government oversight of nanotechnology development [6, 86].

Fundamental to the concerns being raised are questions about whether they can move up the food-chain and reach humans and what exactly nanoparticles do when they enter the human body. Normally harmless conventional compounds can prove to be dangerous on a nanometre scale, as can new nanoparticles being created by scientists [9,19]. Illustrating the fundamental divisions between techno-optimists and sceptics is a debate between Roger Highfield, the science editor of the \emph{Daily Telegraph}, and Bill McKibben, published in October 2003 [87]. For Highfield, ``there is nothing intrinsically good or bad about technology, whether it is GM or nanotechnology, just what we do with it''. McKibben argues instead for foregoing those technologies offering a quantum leap in technological power that ``threaten human meaningÓ and ``human societies''. The debate captures the two ends of the spectrum of views on the place and role of technology in society, offering little hope of resolving such a divide.

Just as research on the ethical implications of nanotechnology is scarce, as discussed for Node 4, reporting on the issue is generally confined to relatively brief statements about funding or legislative measures to deal with ethical issues. Malakoff reported in 
\emph{Science} that the US House and Senate have passed bills requiring studies of ``nanotech's dark side'' [88]. Similarly, a 2003 editorial in \emph{Nature}, while decrying the ``patent nonsense'' of some anti-nanotech campaigns, called for scientists to engage in an ``honest debate'' with the public ``about the potential risks of nanotechnology and how they can be managed'' [89]. Among the few comments addressing at the implications for Third World agricultural production of nano-materials such as nano-textiles, Thomas has called for ``strong global rules to ensure that new technologies are only deployed if the interests of the poor and the vulnerable are protected'' [90].

The paucity of debate and critical analysis on the implications of nanotechnology in the popular media is reflected in the general lack of public awareness of the implications of nanotechnology, as explored next.

\subsection{Node 8: General public}

The general public is, at best, dimly aware of the dimensions of nanotechnology, although the awareness is slowly growing, partly in response to initiatives taken by various governmental and nongovernmental groups [20, 91, 92, 93, 94].

The first representative US national survey of public perceptions of nanotechnology revealed that public awareness of and knowledge about nanotechnology is extremely limited, although ``Americans' initial reaction to nanotechnology is thus far generally positive, probably rooted in a generally positive view of science overall'' [53, p. 395].  Americans do not presume that nanotechnology will lead only to benefits and that there are no potential risks;  and ``the most discouraging aspect to the data is respondents' lack of trust in business leaders to minimize nanotechnology risks to human health'' [53, p. 395]. Of course, anti-nanotechnology activists will likely find this lack of trust to be the most encouraging finding of the survey.

Not surprisingly, although scant, the publicÕs view of nanotechnology probably differs from country to country, depending on national scientific aspirations and climate. This has been exemplified by Gaskell {\em et al.} [95] who analyzed the coverage of nanotechnology in two influential newspapers: the {\em New York Times} and the \emph{Independent} (London). Both newspapers recorded a sixfold increase in the coverage of the risks of nanotechnology from 2000 to 2003, but the US newspaperÕs coverage of the benefits increased almost fivefold in contrast to the twofold increase in the UK newspaper over the same period. This diversity reflects the differences between the cultures of higher and lower technological optimism. Even more importantly, the slant of news coverage not only reflects the current society but also shapes its future.

\section{Conclusion}

	The discussions around nanotechnology epitomize the contemporary processes of making the future present. As Rosenberg and Harding have pointed out, we are ``living through boom times for the future'', an era in which there is a ``remarkable proliferation of words and images about the future'' [96, p. 3]. In such an era, discourses around nanotechnology grapple with the tensions around the boundaries of the real and the hyper-real, development and disaster, human and the posthuman.

Nanotechnology has arisen at a time when biotechnology and information technology are not only highly advanced but continue to grow dramatically. The convergence of these three technologies is fraught with exciting developments for the future of humankind, some with the potential to be extremely beneficial but others that may be extremely deleterious. Government-convened American [27] and British [9] panels have recognized the double-edged nature of this convergence. Even more interesting is the inevitable coupling of these three technologies with cognition science, which has been forecast in movies such as \emph{Gattaca} (1997) and \emph{The Manchurian Candidate} (2004), and has also drawn the attention of the US government [27]. 

Since everyone is in the dark about some or all major aspects of nanotechnology and other emergent technologies, a massive program of general education and information is essential in todayÕs industrial societies, as has been recognized for some years now [2, 8, 97]. A tabulation of the potential social, legal, and ethical implications of nanotechnology over the next three to four decades [98] clearly indicates that formal as well as informal education of technoscientists, politicians, economists, lawyers, social scientists, school teachers, and indeed every citizen is warranted.

Formal education must be imparted at pre-university and university levels to all students, irrespective of areas of specialization, because the societal ramifications of the emergence of nanotechnology and of its convergence with information technology, biotechnology, and cognition science have the potential to be totally transformative---in the same way that the emergence of agriculture must have been at one time.  Informal education too has to be conducted at several levels. At one level, multidisciplinary teams from universities, industry, and the government should conduct seminars for industry leaders, government officials, and legislators. At another level, town councils should fund series of public lectures and panel discussions. Schools, public libraries, and community groups should be provided financial support by government agencies and private foundations to catalyze understanding of all four emergent technologies for all adult learners. Intramural as well as extramural competitions in writing essays, creating items of visual arts, and debates in educational institutions would constitute yet another level of informal education.

It is only through efforts such as these that we will begin to see a fuller range of discourses in Node 8. A better educated public, indeed a focused education of all segments of society in ways suggested above, is likely to facilitate serious and necessary engagement with the issues, concerns and discussions across all nodes. Ongoing critical review of the boundaries that demarcate each node will provide both researchers and policymakers with an informed and nuanced basis for evaluating rapid developments in technoscience more holistically in the future. 

\newpage
\noindent {\bf References}

[1] A. Lakhtakia (ed), Nanometer structures: Theory, modeling, and simulation, SPIE Press, Bellingham, WA, USA, 2004.

[2] R. Merkle, 2001. Nanotechnology: What will it mean? IEEE Spectrum, January, pp. 19-21.

[3] R. Kurzweil, The age of spiritual machines: When computers exceed human intelligence. St Leonards, NSW, Australia: Allen \& Unwin, 1999.

[4] D. Mulhall, Our molecular future: How nanotechnology, robotics, genetics and artificial intelligence will transform our world, Prometheus Books, New York, NY, USA, 2002.

[5] B. McKibben,  Enough: Staying human in an engineered age, Henry Holt, New York, NY, USA, 2003.

[6] B. Joy, The dark side of technology, Vital Speeches of the Day 66 (23) (2000), p. 706.

[7] US National Nanotechnology Initiative, {\sf http://www.nano.gov/html/facts/whatIs Nano.html}

[8] M.C. Roco  and W.S. Bainbridge (eds), Societal implications of nanoscience and nanotechnology, US National Science Foundation, Arlington, VA, USA, 2001.

[9] Royal Society and Royal Academy of Engineering, Nanoscience and nanotechnologies: opportunities and uncertainties, RS Policy document 20/04, RAEng Policy document R2.19, London, United Kingdom, 2004.

[10] J. DiLoreto,  Nanotechnology product stewardship---American Chemistry Council Nanotechnology Panel recognizes value of ASTM Committee E56 efforts,
ASTM International Standardization News  33 (7) (2005), pp. 32-35.

[11] R. Jones, The future of nanotechnology, Physics World 17 (8) (2004), pp. 25-29.

[12] S. Baker and A. Aston, The business of nanotech, Business Week (14 February 2004), pp. 64-71.

[13] C.P. Poole, Jr., and F. J. Owens, Introduction to nanotechnology, Wiley, Hoboken, NJ, USA, 2003.

[14] R.P. Feynman, There's plenty of room at the bottom, Engineering and Science 23 (5) (1960), pp. 22-36.

[15] K.E. Drexler, Molecular engineering: an approach to the development of general capabilities for molecular manipulation, Proc. Nat. Acad. Sci. USA 78 (1981), pp. 5275-5278.

[16] P. Ball,  Nanotechnology in the firing line, {\sf www.nanotech.org} (23 December 2003).

[17] C. Phoenix and E. Drexler, Safe exponential manusfacturing, Nanotechnology 15 (2004), pp. 869-872.

[18] M. Roukes, Plenty of room, indeed. Scientific American (September 2001), pp. 48-57. 

[19] G. Brumfiel, A little knowledge\dots, Nature 424 (2003), pp. 246-248.

[20] {\sf http://www.nanoscience.cam.ac.uk/schools/nano/index.html}

[21] A. Mnyusiwalla, A.S. Daar and P.A. Singer, `Mind the gap': science and ethics in nanotechnology, Nanotechnology  14 (2003), pp. R9-R13.

[22]  J.J. Gilman, Nanotechnology, Mater. Res. Innovat.  5 (2001), pp. 12-14.

[23] R. Roy, Local economy: Nanotech not the answer, Centre Daily Times (State College, PA, USA) (7 September 2004), p. A6.

[24] Y. Gogotsi, How safe are nanotubes and other nanofilaments?, Mater. Res. Innovat. 7 (2003)  pp.192-194.

[25] During a meeting of the Technical Group on Nanotechnology of SPIE---The International Society for Optical Engineering, held on 5 August 2004 in Denver, CO, USA, one of us (AL) presented a summary of Ref. [9] and emphasized the safety and health issue. A large fraction of the audience of technoscientists did not consider the issue as significant, and one person in the audience flatly declared that she was unconcerned about  future detrimental effects to her health.

[26] C.O. Robichaud, D. Tanzil, U. Weilenmann and M.R. Wiesner, Relative risk analysis of several manufactured nanomaterials: An insurance industry context, Environ. Sci. Technol. 10.1021/es0506509 (2005).

[27] M.C. Roco and W.S. Bainbridge, Converging technologies for improving human performance: Nanotechnology, biotechnology, information technology and cognitive science, US National Science Foundation, Arlington, VA, USA, 2002.

[28] D.E. Marlowe,  Nanotechnology and the US Food and Drug Administration,
ASTM International Standardization News 33 (7) (2005), 28-31.

[29] See the letters by N.H. Nevshehir, M. Zelvin, and S. Brauer on p. 20 of the 7 March 2005 issue of {\em Business Week}.

[30] J. Uldrich and D. Newberry, The next big thing is really small: How nanotechnology will change the future of your business, Crown Business, New York, NY, USA, 2003.

[31] Lux Research, Inc., The Nanotech Report 2004. Investment overview and market research for nanotechnology, New York, NY, USA, 2004.

[32] Merrill Lynch issued a press release on April 1, 2004 to announce the creation of a Nanotech Index to track the evolving nanotechnology industry. Available online at 
{\sf http://www.ml.com/index.asp?id=7695$_-$7696$_-$8149$_-$6261$_-$13714$_-$13728}

[33] J. T. Klein, Crossing boundaries: Knowledge, disciplinarities, and interdisciplinarities, Charlottesville, VA, USA, University Press of Virginia, 1996, p. 1.

[34] Nanotechnology: Issues for Business 2005, Brussels, Belgium, 25-26 April (2005). {\sf http://www.euconferences.com/events.asp?ID=2\&Type=Event}

[35] A. Aston, Beaming in on nano gold, Business Week (27 June 2005).

[36] R.W.S. Dunkley, Nanotechnology: social consequences and future implications, Futures 36 (2004), pp. 1129-1132.

[37] P. Picariello, New ASTM International Committee E56 on Nanotechnology,
ASTM International Standardization News 33 (7) (2005), p. 27.

[38] H. Shindo, Nanotechnology standardization in Japan, ASTM International Standardization News 33 (7) (2005), pp. 36-39.

[39] F. A. Fiedler and G. H. Reynolds, Legal Problems of Nanotechnology: An Overview, Southern California Interdisciplinary Law Journal, 3 (1994), pp. 593-630.

[40] P. C. Lin-Easton, It's Time for Environmentalists to Think Small--Real Small: A Call for the Involvement of Environmental Lawyers in Developing Precautionary Policies for Molecular Nanotechnology, Georgetown International Environmental Law Review, 14 (2001), pp. 107-134.

[41] G. H. Reynolds, Environmental Regulation of Nanotechnology: Some Preliminary Observations, Environmental Law Reporter (2001), pp. 10681-10866.

[42] B. Newberger, Intellectual Property and Nanotechnology, Texas Intellectual Property Law Journal, 11 (2003), pp. 649-657.

[43] A. P. Halluin and L. P. Westin, Nanotechnology: The Importance of Intellectual Property Rights in an Emerging Technology, Journal of the Patent and Trademark Office Society, 86 (2004), pp. 220-236.

[44] I. Kerr and G. Bassi, Not That Much Room?  Nanotechnology, Networks and the Politics of Dancing, Health Law Journal, 12 (2004), pp. 103-123.

[45] R. H. Smith, Social, Ethical, and Legal Implications of Nanotechnology, Societal Implications of Nanotechnology, National Science Foundation, Washington, DC, 2001, pp. 203-211.

[46] W. D. Kay and C. J. Bosso, A Nanotech Velvet Revolution?  Issues for Social Science Inquiry, STEP Ahead, 3 (2005), pp. 2-3.

[47] J. Altmann, Military Uses of Nanotechnology: Perspectives and Concerns, Security Dialogue 35 (2004), pp. 61-79.

[48] P. B. Thompson, Social Acceptance of Nanotechnology, Societal Implications of Nanoscience and Nanotechnology, National Science Foundation, Washington, DC, 2001, pp. 198-202.

[49] V. Weil, Ethical Issues in Nanotechnology, In: M. C. Roco and W. S. Bainbridge, eds., Societal Implications of Nanoscience and Nanotechnology, National Science Foundation, Washington, DC, 2001, pp. 193-198.

[50] R. M. Satava, Disruptive Visions: Moral and Ethical Challenges from Advanced Technology and Issues for the New Generation of Surgeons, Surgical Endoscopy 16 (2002), pp. 1403-1408.

[51] M. C. Roco, Broader Societal Issues of Nanotechnology, Journal of Nanoparticle Research 5 (2003), pp. 181-189.

[52] W. S. Bainbridge, Sociocultural Meanings of Nanotechnology: Research Methodologies, Journal of Nanoparticle Research 6 (2004), pp. 285-299.

[53] M. D. Cobb and J. Macoubrie, Public Perceptions about Nanotechnology: Risks, benefits and Trust, Journal of Nanoparticle Research 6 (2004), pp. 395-405.

[54] T. Braun, Nanoscience and Nanotechnology in the Balance, Scientometrics 38 (1997), pp. 321-325.

[55] M. Meyer, Nanotechnology: Interdisciplinarity, Patterns of Collaboration and Differences in Application, Scientometrics 42 (1998), pp. 195-205.

[56] H. Eto, Interdisciplinary information Input and Output of a Nano-Technology Project, Scientometrics 58 (2003), pp. 5-33.

[57] J. Schummer, Multidiciplinarity, Interdisciplinarity, and Patterns of Research Collaboration in Nanoscience and Nanotechnology, Scientometrics 59 (2004), pp. 425-465.

[58] A. Hullmann, Publications and Patents in Nanotechnology: An Overview of Previous Studies and the State of the Art, Scientometrics 58 (2003), pp. 507-527.

[59] M. F. Bendle, Teleportation, Cyborgs and the Posthuman Ideology, Social Semiotics 12 (2002), pp. 45-62.

[60] C. Elliott, Humanity 2.0, Wilson Quarterly 27 (2003), pp. 13-20.

[61] J. Gimzewski and V. Vesna, The nanoneme syndrome: Blurring of fact and fiction in the construction of a new science. Technoetic Arts 1 (1) (2003), pp. 7-24.

[62] K.E. Drexler, Engines of creation: The coming era of nanotechnology,  Anchor-Doubleday, New York, NY, USA, 1986.

[63] M. Crichton, Prey, HarperCollins Publishers, New York, 2002.

[64] N. Stephenson, The diamond age, or a young lady's illustrated
primer, Bantam Books, New York, 1995. 

[65] C. Milburn, Nanotechnology in the age of posthuman engineering: Science fiction as science, Configurations 10 (2) (2002), pp. 261-295.

[66] J. Baudrillard, The spirit of terrorism, Verso, London, UK, 2002.

[67] F. Fukuyama,   Our posthuman future: Consequences of the biotechnology revolution, Farrar, Straus, \& Giroux, New York, NY, USA, 2002.

[68] See the blurb on the jacket of Ref. [5].

[69] ETC; Intermediate Technology Development Group; Research Foundation for Science, Technology and Ecology headed by Vandana Shiva; Greenpeace; and the  International Centre for Bioethics, Culture and Disability are excellent examples.

[70] B. Feder, Prince's technology qualms create a stir in Britain, New York Times (19 May 2003), p. C3.

[71] Z. Goldsmith, Editorial, The Ecologist 34 (7) (2004), p. 4.

[72] ETC, Down on the farm: The impact of nano-scale technologies on
food and agriculture, November 2004,
\noindent{\sf http://www.etcgroup.org/documents/ETC$_-$DOTFarm2004. pdf }

[73] ETC, NanotechÕs ``second nature'' patents: Implications for the
global south, Communiques No. 87 \& 88, March/April and May/June 2005, 
\noindent{\sf http://www.etcgroup. org/documents/Com8788SpecialPNanoMar-Jun05ENG.pdf }

[74] ETC, NanoGeoPolitics: ETC Group surveys the political landscape,
Special Communique No. 89, July 2005, 
\noindent{\sf http://www.etcgroup.org/documents/Com89Special NanoPoliticsJul05ENG.pdf }

[75] B. Feder, Research shows hazards in tiny particles, New York Times
(14 April 2003), p. C8.

[76] A.H. Arnall, Future technologies, today's choices, Greenpeace Environmental Trust, London, UK, 2003.

[77] A. Regalado, Greenpeace warns of pollutants from nanotechnology, Wall Street Journal (25 July 2003), p. B1.

[78] {\sf http://www.etcgroup.org/key$_-$defs.asp}

[79] J. Thomas, Nanowatch,  The Ecologist 34 (4) (2004), p. 13.

[80] J. Thomas,  Future perfect?, The Ecologist 33 (4) (2003), p. 32.

[81] J. Thomas, Little BrotherÕs watching you, The Ecologist 33 (9) (2003), p.16.

[82] P. Montague, Welcome to NanoWorld: Nanotechnology and the precautionary principle imperative, Multinational Monitor 25 (9) (2004), p. 16-19.

[83] B. Schecter, They've seen the future and intend to live it, New York Times (16 July  2002), p. F4.

[84] L. Gomes, Boom town: Beware of Valley's new fad: Nanotechnology, Wall Street Journal (6 May 2002), p. B1.

 [85] B. Feder, Nanotechnology has arrived; a serious opposition is forming, New York Times (19 August 2002), p. C3.

[86] R. Service, Is nanotechnology dangerous?, Science 290 (2000), p. 1526. 

[87] R. Highfield and B. McKibben, Technology: Will it save us or kill us?, The Ecologist 33 (10) (2003), pp. 40-42 \& 57.

[88] D. Malakoff, Congress wants studies of NanotechÕs ``dark side'', Science 301 (2003), p. 27.

[89] Don't believe the hype, Nature 424 (2003), p. 237.

[90] J. Thomas, Atomising Third World economies, The Ecologist 34 (8) (2004), p. 18.

[91] A. Lakhtakia, Nanotechnology offers promise on a smaller scale, Centre Daily Times (State College, PA, USA) (31 January 2005), p. A6.

[92] {\sf http://www.nanojury.org/}

[93] {\sf http://www.uea.ac.uk/env/cer/index.html}

[94] {\sf http://www.nano-and-society.org/}

[95] G. Gaskell, T.T. Eyck, J. Jackson and G. Veltri, Public attitudes to nanotechnology in Europe and the United States, Nature Materials 3 (2004), p. 496.

[96] D. Rosenberg and S. Harding, Introduction: Histories of the future. In: D. Rosenberg and S. Harding (Eds.), Histories of the future. Duke University Press, Durham, NC, USA, 2005.

[97] A. Lakhtakia, Taking nanotechnology to schools,
\noindent {\sf http://www.arXiv.org/physics/ 0505007}

[98] R.H. Smith, Social, ethical, and legal implications of nanotechnology, In: M. C. Roco and W. S. Bainbridge, eds., Societal Implications of Nanoscience and Nanotechnology, National Science Foundation, Washington, DC, 2001.
pp. 203-211.

\end{document}